\documentclass[
reprint,
longbibliography,
amsmath, 
amssymb, 
aps, 
superscriptaddress,
prx,
]
{revtex4-2}
\usepackage{graphicx}
\usepackage{dcolumn}
\usepackage{bm}
\usepackage{braket} 
\usepackage{hyperref}
\usepackage{subfigure} 
\usepackage{nccmath}
\usepackage{amsthm} 
\usepackage{mathrsfs} 
\usepackage{xfrac} 
\usepackage{pifont}
\usepackage[export]{adjustbox}
\usepackage{bbm} 
\usepackage[utf8]{inputenc} 
\usepackage[normalem]{ulem}
\usepackage[dvipsnames,svgnames,table]{xcolor}
\usepackage[sort&compress]{natbib}
\usepackage{color}
\usepackage{longtable}
 
\usepackage{commath}
\usepackage{multirow}
\usepackage{array}

\newcommand{\epsa}{\varepsilon_{\mathrm{acc}}}
\newcommand{\povl}{f_{\mathrm{ovl}}}

\newcommand{\lambdaeff}{\lambda_{\mathrm{eff}}}
\newcommand{\pL}{p_{L}}

\begin{document}

\title{Evaluating the performance of QEC primitives on quantum processors at large width and depth}

  \author{J. A. Monta\~nez-Barrera}
    \altaffiliation{j.montanez-barrera@fz-juelich.de}
	\affiliation{Jülich Supercomputing Centre, Forschungszentrum Jülich, D-52425 Jülich, Germany}

    \author{Kristel Michielsen}
	\affiliation{Jülich Supercomputing Centre, Forschungszentrum Jülich, D-52425 Jülich, Germany}
	\affiliation{Department of Computer Science, University of Cologne, 50931 Cologne, Germany }

\begin{abstract}
Quantum error correction (QEC) relies on repeated parity extraction, mid-circuit measurement (MCM), reset, feed-forward, and scheduling, yet these primitives are usually assessed either in isolation or through resource-demanding experiments. We introduce a few-sample benchmark of QEC-relevant primitives based on an MCM implementation of the quantum approximate optimization algorithm (QAOA) with a fixed set of linear parameters (LR-QAOA). For a chosen code, the QAOA Hamiltonian is constructed from its check structure, such that the resulting circuit mimics the syndrome-extraction connectivity and MCM pattern while producing a direct algorithmic signal, the approximation-ratio $r$. The LR-QAOA depth, defined by the number of QAOA layers, plays a role analogous to the number of repeated syndrome-extraction rounds in a QEC experiment. From QPU executions, the decay of $r$ with depth defines an effective hardware error, which we map to an equivalent two-qubit depolarizing rate $\lambda_{\mathrm{eff}}$. We compare direct and MCM-mediated implementations across $10$ QPUs from IBM, IQM, and Quantinuum, with circuits containing up to $2950$ MCM operations. On Quantinuum's \texttt{Helios-1} and \texttt{H2-1}, we run code-structured LR-QAOA benchmarks for surface-code, triangular color-code, and bivariate-bicycle qLDPC Hamiltonians up to $81$, $91$, and $48$ data qubits, respectively, using up to $480$ MCM operations. On IBM \texttt{ibm\_phoenix}, we implement the surface-code structure and compare the LR-QAOA response with logical-memory experiments, observing a correlation between the benchmark and the logical performance across different regions of the QPU. Its construction and low resource requirements provide a practical benchmark for comparing hardware generations and QEC implementations before full logical-memory experiments are performed.

\begin{description}
	\vspace{0.2cm}
	\item[Keywords] QEC Benchmarking, mid-circuit measurement, Quantinuum Helios-1, IBM Nighthawk-r2, IQM Emerald, color-code, surface-code.
\end{description}
\end{abstract}

\maketitle
\section{Introduction}\label{Sec:Introduction}

Although many promising quantum error-correcting codes are known, realizing fault tolerance in hardware requires reliable implementation and validation of the repeated, measurement-driven syndrome-extraction circuits on which those codes depend. Each round of stabilizer readout uses the same ingredients that are hardest to characterize in isolation: ancilla-data entangling gates, mid-circuit measurement (MCM), reset, feed-forward, and idling of spectator qubits.

Surface-code experiments have progressed from repeated syndrome extraction to below-threshold logical memories with real-time decoding \cite{QECGoogle2024}, while high-rate low-density-parity-check (qLDPC) codes have become a leading route to reduce overhead \cite{Bravyi_2024,Ye_2025}. Recent trapped-ion demonstrations further show that qLDPC, topological, and concatenated codes with very different connectivity requirements can be implemented on the same device, with small qLDPC instances reaching breakeven performance \cite{tham2026breakevendemonstrationquantumlowdensity}. The $[[7,1,3]]$ Steane code, equivalent to a distance-$3$ triangular color code, has also been used in error corrected computations involving multiple logical qubits \cite{perlin2026}. 

As the roadmaps of different quantum vendors increasingly target QEC demonstrations \cite{ibm_quantum_roadmap,quantinuum_roadmap_2024,ionq_roadmap,iqm_roadmap}, it becomes important to evaluate quantum processing units (QPUs) in terms of their raw QEC-relevant dynamic-circuit performance in a resource-efficient manner. This is particularly relevant for independent institutions seeking to assess whether a given technology is progressing consistently with its roadmap, since they may neither have the experimental expertise nor be able to dedicate the substantial computational and sampling resources required for full decoded logical experiments, where multiple additional factors can also influence the observed performance.

The current characterization toolbox spans several complementary levels. Component-level methods such as randomized benchmarking and gate-set tomography quantify errors associated with gates or small sets of operations \cite{Knill_2008,Magesan_2012,Nielsen_2021}. Dedicated MCM benchmarks extend this characterization to dynamic-circuit primitives by identifying measurement-induced errors on both the measured qubit and unmeasured spectators, including crosstalk and entangling errors generated by the measurement process itself \cite{Govia_2023, hothem2024measuringerrorratesmidcircuit}. These approaches provide detailed information about individual hardware primitives, but do not by themselves capture how those errors combine within a complete QEC-relevant circuit structure.

More recent dynamic-circuit benchmarking protocols extend this direction by interleaving measurement-and-feed-forward blocks within randomized benchmarking sequences \cite{Shirizly_2025}, and by using application-level dynamic circuits, structural features, and scalable fidelity scores to predict hardware performance across devices and calibration cycles \cite{shirgure2026characterizingbenchmarkingdynamicquantum}. More directly in the QEC setting, scalable noise-characterization protocols have been proposed for syndrome-extraction circuits \cite{Hockings_2025}. These methods aim to reconstruct detailed gate-level noise information and have been numerically demonstrated for surface-code circuits with more than $10^3$ qubits. At the other end of the QEC stack, rare-event sampling techniques have been proposed to estimate logical failure rates when error correction suppresses them to levels that are difficult to access through direct sampling \cite{beverland2025}. 

Here, we target an intermediate regime that can provide a meaningful signal of how quantum hardware is evolving toward QEC without requiring extensive QPU resources. The benchmark is designed to be practical to implement and flexible enough to adapt to different QEC schemes. Rather than reconstructing the underlying noise model or estimating rare logical failures, we use a low-sampling algorithmic observable to evaluate the aggregate performance of QEC-relevant dynamic circuits. The quantum approximate optimization algorithm (QAOA) \cite{farhi2014} provides a suitable foundation for such a benchmark.

We exploit four characteristics of the algorithm. First, the multibody terms of a higher-order Ising Hamiltonian can be implemented using MCMs, allowing the impact of MCM operations to be quantified. Second, the structure of a QEC code can be mapped onto the QAOA problem Hamiltonian, yielding a higher-order Ising model whose interaction structure follows that of the QEC code. Third, increasing the algorithmic depth is analogous to applying subsequent rounds of syndrome extraction. In the ideal case, the quality of the QAOA solution improves with depth, whereas on real hardware the accumulation of noise progressively degrades this signal. In this way, the resulting circuit reproduces the structure and resource requirements of a syndrome-extraction circuit while providing an algorithmic signal that is directly sensitive to accumulated hardware errors. Finally, noisy QAOA signals can be distinguished from random sampling, and differences between error rates can be resolved, using relatively few samples.

The resulting protocol serves two complementary purposes. As a hardware benchmark, it quantifies the degradation introduced by MCM in dynamic circuits; here, we demonstrate this use on IBM, IQM, and Quantinuum devices. As a QEC-readiness benchmark, the circuit can be built directly from the stabilizer checks of any QEC code, reproducing its syndrome-extraction connectivity and measurement structure. We illustrate this flexibility using surface, color, and qLDPC codes. LR-QAOA thus provides an intermediate benchmark between device calibration and logical performance, enabling hardware platforms, code constructions, and syndrome-extraction schedules to be tested before undertaking a decoded-memory study.

The paper is organized as follows. Section~\ref{Sec:Methods} details the LR-QAOA schedule, the Hamiltonian constructions, the overlap and effective-rate definitions, the logical mapping, and the experimental setup. Section~\ref{Sec:Results} presents the results: the effective cost of the MCM primitive across platforms (Sec.~\ref{Sec:res-mcm}), the LR-QAOA QEC structure evaluation (Sec.~\ref{Sec:res-structure}), the shot-cost to distinguish noisy devices (Sec.~\ref{Sec:sampling-budget}), and a memory experiment versus the LR-QAOA estimation (Sec.~\ref{Sec:qec-vs-qaoa}) Finally, Section~\ref{Sec:Conclusions} presents the conclusion.

\section{Methods}\label{Sec:Methods}
\begin{figure*}[!tbh]
\centering
\includegraphics[width=16cm]{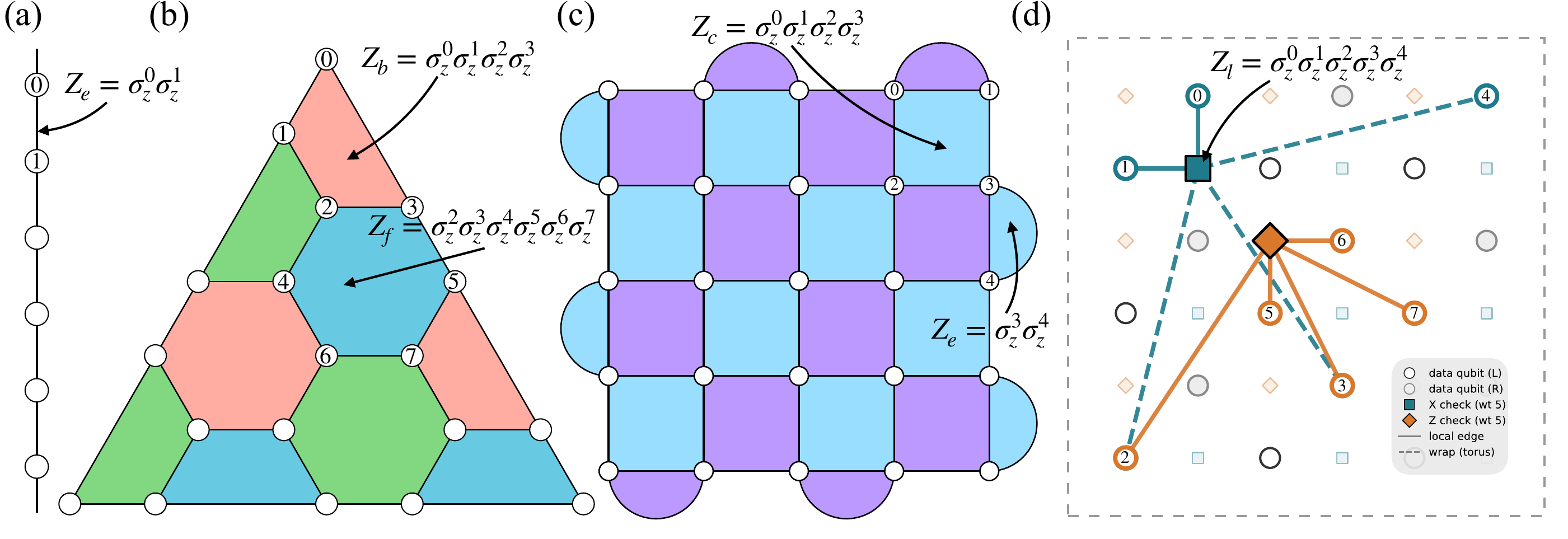}
\caption{\label{Fig:Hcs} Hamiltonians generation. (a) 1D-Chain (b) $d=5$ color-code with $Z_b$ 4-body terms $\mathcal F_{bdry}^{(4)}$ and $Z_f$ 6-body terms $\mathcal F_{bulk}^{(6)}$ (c) $d=5$ surface-code with 2-body terms $\mathcal C_{bdry}^{(2)}$ and $Z_f$ 4-body terms $\mathcal C_{bulk}^{(4)}$. (d) BB$[[18,4,3]]$ bivariate-bicycle qLDPC code; its independent $X$- and $Z$-checks all have the same weight, so they form a single set $\mathcal C^{(5)}$ of 5-body $Z$-terms $Z_l$. The long-range edges are the couplings that wrap around the $\ell\times m$ torus.}
\end{figure*}
\subsection{The linear ramp angle schedule}\label{sec:methods-basis}

To determine the QAOA angles, we use LR-QAOA, a variant in which the cost and mixer angles follow linear schedules specified by only two variational parameters, corresponding to the slopes of these schedules \cite{Montanez_Barrera_2025_lrqaoa}. It can be viewed as a first-order Trotterized approximation to adiabatic quantum evolution~\cite{farhi2000quantum}. The algorithm alternates between applying a problem Hamiltonian $H_C$, whose ground state encodes the desired solution, and a mixer Hamiltonian $H_B$, whose ground state initializes the evolution. Over $p$ layers, the average energy is progressively reduced, increasing the probability of sampling high-quality solutions. This energy decrease, which has been shown to grow rapidly with circuit depth \cite{montanezbarrera2025benchmarking, Montanez_Barrera_2025_lrqaoa}, provides a useful signal for hardware benchmarking. Improvements in this signal can therefore be interpreted as evidence of progress across hardware generations. These properties have also motivated the adoption of LR-QAOA in benchmarking frameworks such as Metriq-Gym and IonQ's application-oriented benchmarking suite \cite{cosentino2026metriqcollaborativeplatformbenchmarking,aboumrad2026measuringmattersscalableframework}.

The LR-QAOA angle
schedule follows a linear ramp:
\begin{equation}\label{Eq:ramps}
    \gamma_l = \frac{l\,\Delta\gamma}{p},
    \qquad
    \beta_l  = \frac{(p+1-l)\,\Delta\beta}{p},
    \qquad l = 1, \ldots, p,
\end{equation}
so that $\gamma_l$ increases monotonically from $\Delta\gamma/p$ to $\Delta\gamma$,
while $\beta_l$ decreases monotonically from $\Delta\beta$ to $\Delta\beta/p$. Each layer applies a distinct unitary
\begin{equation}
    U_l = U_B(\beta_l)\,U_C(\gamma_l)
        = e^{-i\beta_l H_B}\,e^{-i\gamma_l H_C}.
    \label{eq:layer}
\end{equation}
where $H_C$ is the cost Hamiltonian and the circuit is initialized in the uniform superposition state $\ket{+}^{\otimes N_q}$ for $N_q$ qubits. In all benchmarks considered here, $H_C$ is expressed as a sum of Pauli-$Z$ interaction terms. For a subset of qubits $s$, we define
\begin{equation}\label{Eq:z_terms}
    Z_s = \prod_{i\in s} \sigma_z^{(i)} ,
\end{equation}
so that the support $s$ determines the locality of each interaction and $\sigma_z^i$ denotes the Pauli-$z$ operator acting on qubit $i$, which represents a vertex of the graph.

For the 1D chain, Fig.~\ref{Fig:Hcs}(a), the supports are the nearest-neighbor edges
$e\in\mathcal E_{1D}$, giving
\begin{equation}
    H^{\mathrm{1D}}_C
    =
    \sum_{e\in \mathcal E_{1D}} Z_e ,
\end{equation}
with $Z_e=\sigma_z^{(i)}\sigma_z^{(j)}$ for $e=(i,j)$.

For the triangular color-code Hamiltonian, Fig.~\ref{Fig:Hcs}(b), the supports are the faces of the triangular lattice. The bulk faces give six-body interactions, while the boundary faces give four-body interactions,
\begin{equation}\label{Eq:CC}
    H^{\mathrm{CC}}_{C}(d)
    =
    \sum_{f\in \mathcal F_{\mathrm{bulk}}^{(6)}(d)} Z_f
    +
    \sum_{b\in \mathcal F_{\mathrm{bdry}}^{(4)}(d)} Z_b .
\end{equation}
Here, $\mathcal F_{\mathrm{bulk}}^{(6)}(d)$ and
$\mathcal F_{\mathrm{bdry}}^{(4)}(d)$ denote the corresponding sets of six-body and four-body face terms.

For the surface-code Hamiltonian, Fig.~\ref{Fig:Hcs}(c), the supports are the plaquette checks of the lattice. These consist of four-body bulk checks and two-body boundary checks,
\begin{equation}
    H^{\mathrm{SC}}_{C}(d)
    =
    \sum_{c\in \mathcal C_{\mathrm{bulk}}^{(4)}(d)} Z_c
    +
    \sum_{e\in \mathcal C_{\mathrm{bdry}}^{(2)}(d)} Z_e .
\end{equation}
Here, $\mathcal C_{\mathrm{bulk}}^{(4)}(d)$ and
$\mathcal C_{\mathrm{bdry}}^{(2)}(d)$ denote the sets of four-body and two-body checks, respectively at some distance $d$.

For the bivariate-bicycle (BB) qLDPC Hamiltonian, Fig.~\ref{Fig:Hcs}(d), the $n=2\ell m$ data qubits split into two sublattices ($L,R$) placed on an $\ell\times m$ torus. Unlike the color and surface codes, where the sum is split because the bulk and boundary supports have different body number, here the independent (mod $2$) $X$- and $Z$-checks all share the same weight $w$, so they combine into a single set
$\mathcal C^{(w)}$ of $w$-body terms,
\begin{equation}
    H^{\mathrm{qLDPC}}_{C}
    =
    \sum_{c\in \mathcal C^{(w)}} Z_l ,
\end{equation}
with $|\mathcal C^{(w)}|=n-k$ for $n$ physical qubits encoding $k$ logical qubits. Every term is a translation of the same weight-$w$ stencil (the code is quasi-cyclic), so a single support generates the whole family by lattice translation. The instance in Fig.~\ref{Fig:Hcs}(d) is the BB$[[18,4,3]]$ code, with $\ell=m=3$ ($n=18$ data qubits, giving $|\mathcal C^{(5)}|=14$ five-body terms $Z_l$; the terms that wrap across the $\ell\times m$ torus are the long-range couplings characteristic of qLDPC codes. This same bivariate-bicycle code family was used in the recent below-breakeven memory demonstration on a trapped-ion processor by IonQ~\cite{tham2026breakevendemonstrationquantumlowdensity}.

The same construction can be implemented in the conjugate basis. Applying $H$ to every data qubit maps $H_C \mapsto H^{\otimes N_q} H_C H^{\otimes N_q}$, such that each check becomes $\bigotimes_{q\in c} X_q$. Correspondingly, the mixer becomes $H_B = \sum_i Z_i$, the initial state is $|0\rangle^{\otimes N_q}$, the measure-and-correct gadget applies $X$ corrections through ancilla-controlled CNOTs, and the data register is measured in the $X$ basis. Since the two constructions are related by a unitary conjugation, they have identical noiseless energy distributions and therefore the same $r$ in the ideal case. Thus, $r$ can be directly compared between the two implementations, while the conjugate-basis construction allows us to directly probe the effect of dephasing on performance and represent better the X-checks construction of the QEC code. We use this methodology below when comparing against memory experiments.

We use the approximation ratio as the main performance metric. Given a measured
bitstring $z$, its energy is evaluated as
\begin{equation}
    E(z) = \langle z | H_C | z \rangle ,
\end{equation}
For a sampled output distribution $P(z)$,
the average energy is then
\begin{equation}
    \langle E \rangle = \sum_z P(z) E(z).
\end{equation}
The approximation ratio is defined as
\begin{equation}
    r =
    \frac{\langle E\rangle - E_{\max}}
         {E_{\min} - E_{\max}},
    \label{eq:approx_ratio}
\end{equation}
where $E_{\min}$ and $E_{\max}$ are the minimum and maximum energies over all
bitstrings, respectively. With this convention, $r=1$ corresponds to sampling
only optimal solutions, while $r=0$ corresponds to the worst-cost configuration.

In addition to the approximation ratio, for small circuits we also report the
probability of success. This quantity is defined as the probability of sampling
an optimal bitstring,
\begin{equation}
    P_{\mathrm{succ}}
    =
    \sum_{z \in z_{\mathrm{opt}}} P(z),
    \label{eq:prob_success}
\end{equation}
where $z_{\mathrm{opt}}=\{z : E(z)=E_{\min}\}$ is the set of optimal
bitstrings.

\subsection{The signal overlap}\label{Sec:noise}

LR-QAOA exhibits a characteristic noise response that can be exploited to define a depolarizing-noise-equivalent error for the noise perceived on quantum hardware on a simple 1D-Chain experiment. The goal is to map the algorithmic degradation observed on a QPU to an equivalent depolarizing error rate, providing a common metric to quantitatively compare circuits executed with a direct implementation, based only on CNOT gates, and with an implementation in which MCM operations mediate the interactions between qubits. The difference between these two effective error rates then provides a measure of the additional hardware cost associated with introducing MCM operations in increasingly large chains of qubits.

For an observable $\mathcal O$, a density operator $\rho_p^{(\lambda)}$ obtained at circuit depth $p$ and depolarizing noise strength $\lambda$, we define the normalized overlap as
\begin{equation}
\povl^{\mathcal O}(p,\lambda) = 
\frac{\mathcal O(\rho_{p}^{(\lambda)})-\mathcal O_{\mathrm{rand}}}
{\mathcal O(\rho_{p}^{(0)})-\mathcal O_{\mathrm{rand}}}.
\label{eq:normalized_overlap}
\end{equation}
Here, $\mathcal O(\rho_{p}^{(\lambda)})$ is the noisy value obtained either from the QPU or from a depolarizing-noise simulation, $\mathcal O(\rho_{p}^{(0)})$ is the corresponding noiseless value, obtained from state vector or MPS simulation, and $\mathcal O_{\mathrm{rand}}$ is the random-sampling limit, estimated using a random sampler. With this definition, $\povl^{\mathcal O}=1$ corresponds to the noiseless result, whereas $\povl^{\mathcal O}=0$ indicates that the observable has decayed to the random-sampling limit.

To characterize this decay, we perform two-qubit depolarizing-noise simulations and define the accumulated error as
\begin{equation}
\epsa = N_{\mathrm{edges}}p\lambda,
\label{eq:eps_acc}
\end{equation}
where $N_{\mathrm{edges}}$ is the number of interactions per LR-QAOA layer and $\lambda$ is the depolarizing-noise-equivalent error associated with each edge. For a fixed problem size and observable, the normalized overlap approximately collapses onto a single exponential,
\begin{equation}
\povl^{\mathcal O}(\epsa)
\approx
2^{-\kappa_{\mathcal O}\epsa},
\label{eq:depol_collapse}
\end{equation}
where $\kappa_{\mathcal O}$ quantifies the susceptibility of the observable to accumulated noise. From the depolarizing-noise simulations with 3 to 15 qubits, we find that the susceptibility of the approximation ratio depends on the number of qubits as
\begin{align}
\kappa_{r}(N_q)
&\simeq \frac{\kappa_{0}}{N_q},
&& \kappa_{0}\simeq 3.69,
\label{eq:kappa_ar}
\end{align}
whereas for the probability of success it is approximately independent of system size,
\begin{align}
\kappa_{P_{\mathrm{succ}}}
&\simeq \kappa_{P},
&& \kappa_{P}\simeq 1.
\label{eq:kappa_prob}
\end{align}

This depolarizing-noise response provides a reference against which the hardware results can be compared. For a QPU, we define an effective depolarizing error $\lambda_{\mathrm{eff}}$ such that the corresponding depolarizing-noise model reproduces the algorithmic degradation observed on the device,
\begin{equation}
f_{\mathrm{ovl},\mathrm{hw}}^{r}
\approx
2^{-\kappa_{r}N_{\mathrm{edges}}p\lambda_{\mathrm{eff}}}.
\label{eq:hardware_model}
\end{equation}
Importantly, $\lambda_{\mathrm{eff}}$ should not be interpreted as the physical error rate of a single hardware operation. Rather, it represents the depolarizing error per edge that would produce the same degradation of the LR-QAOA signal as the combined noise sources present in the hardware implementation.

We extract this quantity independently for the direct and MCM implementations, obtaining $\lambda_{\mathrm{direct}}$ and $\lambda_{\mathrm{mcm}}$, respectively. For the 1D LR-QAOA circuits studied here, the corresponding accumulated errors are therefore
\begin{equation}
\epsa^{\mathrm{direct(mcm)}}=N_{\mathrm{edges}}p\lambda_{\mathrm{direct(mcm)}},
\label{eq:eps_acc_direct}
\end{equation}
with $N_{\mathrm{edges}}=N_q-1$. The direct implementation contains $N_{\mathrm{2q}}=2N_{\mathrm{edges}}$ two-qubit gates per layer. The MCM implementation contains the same number of two-qubit gates, together with $N_{\mathrm{edges}}$ MCM operations per layer. Consequently, the difference between $\lambda_{\mathrm{mcm}}$ and $\lambda_{\mathrm{direct}}$ quantifies the additional algorithmic degradation introduced by the MCM-mediated implementation, including the measurement itself and the hardware overhead associated with its execution.

\subsection{How many shots to rank two devices}\label{Sec:res-rank}
The perceived rate $\lambdaeff$ is meant to be compared across processors, so we estimate how many shots does it take to decide, from the measured approximation ratio alone, which of two devices is better. For two devices with perceived two-qubit error rates $\lambda_A>\lambda_B$, the overlap model gives the mean approximation ratio
\begin{equation}
  r(\lambda)
  =
  r_{\mathrm{rand}}
  +
  (r_{\mathrm{ideal}}-r_{\mathrm{rand}}) \,
  2^{-\kappa N_g\lambda},
  \label{eq:ranking_mean}
\end{equation}
where $N_g$ is the number of gates with perceived depolarizing noise $\lambda$. Thus the expected separation is
$\Delta r=r(\lambda_B)-r(\lambda_A)>0$.  If each device is sampled with $S$ shots and the two sample means are approximately Gaussian, then
\begin{equation}
  z(S)
  =
  \frac{\Delta r\,\sqrt{S}}
       {\sqrt{\sigma_A^2+\sigma_B^2}},
  \qquad
  P(\hat r_B>\hat r_A)=\Phi[z(S)] ,
  \label{eq:ranking_z}
\end{equation}
where $\sigma_A^2$ and $\sigma_B^2$ are the per-shot variances of the measured approximation ratio on the two devices, and $\Phi$ denotes the cumulative
distribution function of a standard normal random variable.  The number of shots per device needed to reach a target one-sided confidence $z$ is therefore
\begin{equation}
  S^\ast=\frac{z^2(\sigma_A^2+\sigma_B^2)}{\Delta r^2}.
  \label{eq:sstar}
\end{equation}
This test uses only the measured approximation ratio; the noiseless (ideal) value is needed only to interpret the operating point.

The optimal depth follows from the same decay law.  Giving $\lambda_A=\lambda$ and $\lambda_B=(1-\delta)\lambda$, with $\delta=(\lambda_A-\lambda_B)/\lambda_A$. For $x=\kappa N_g\lambda$, the exact separation predicted by Eq.~\eqref{eq:ranking_mean} is
\begin{equation}
  \Delta r
  =
  (r_{\mathrm{ideal}}-r_{\mathrm{rand}}) \,2^{-x}\!\left(2^{\delta x}-1\right).
  \label{eq:dmu_exact}
\end{equation}
For a small relative gap, $\delta x\ll1$, this becomes
$\Delta\mu\simeq \Delta r\,(\ln2)\,\delta\,x\,2^{-x}$. The distinguishing power is therefore maximal at
$x^\ast=1/\ln2$, i.e. at $f_{\mathrm{ovl}}=2^{-x^\ast}=1/e\approx0.37$, the transition region between the
ideal and random limits.  If the gate count grows linearly with LR-QAOA depth, $N_g(p)\simeq pN_{2q}$, this operating point is reached at
\begin{equation}
  p^\ast\simeq \frac{1}{\ln2\,\kappa N_{2q}\lambda_A}.
  \label{eq:pstar}
\end{equation}
At this point, $\Delta r_{\max}\simeq (r_{\mathrm{ideal}}-r_{\mathrm{rand}}) \,\delta/e$, and
\begin{equation}
  S^\ast_{\min}\simeq\frac{z^2 e^2(\sigma_A^2+\sigma_B^2)}
  {(r_{\mathrm{ideal}}-r_{\mathrm{rand}})^2\,\delta^2},
  \label{eq:sstar_min}
\end{equation}
with the variances evaluated near the optimal operating point.  The shot cost is controlled by the relative gap $\delta$, and scales as $1/\delta^2$. Reducing the absolute error rate does not by itself make ranking harder; it moves the optimal benchmark to larger depth, $p^\ast\propto1/\lambda_A$.

Similar to what has been observed for square-lattice, heavy-hex and fully connected structures in \cite{montanezbarrera2025benchmarking}, the statistical error on the measured ratio scales as $\varepsilon=O\bigl((M|E|)^{-1/2}\bigr)$ for $|E|$ hyperedges and $M$ shots.

\subsection{LR-QAOA experimental setup}

To evaluate the performance of the \emph{MCM} primitive on real hardware, we use a dynamic-circuit representation of the $Zs$ Hamiltonian terms in Eq.~\ref{Eq:z_terms}, which requires an ancillary qubit to mediate the interaction. We compare this approach with a \emph{direct} implementation, where the interaction is implemented only through CNOT gates acting on the corresponding data qubits.

The first problem we consider is a 1D chain of qubits, which represents the simplest QAOA instance. We solve this problem using both a direct and MCM implementations with the corresponding circuits shown in Fig.~\ref{Fig:experimental-setup}(a). Since different quantum computing hardware platforms are at different stages of maturity with respect to MCM operations, this initial test provides a useful baseline to assess the performance of a minimal system consisting of two data qubits and a single ancillary qubit mediating the interaction.

In the case of the different QEC-code-inspired Hamiltonians, the Hamiltonian structure can also be implemented either directly or through an MCM-mediated construction. For instance, Fig.~\ref{Fig:experimental-setup}(b) shows the implementation for the SC case. The corresponding resource counts for the three cases considered in this work are summarized in Table~\ref{Tab:gate-counts}.

\begin{table}[t]
    \centering
    \caption{Number of two-qubit gates and MCM operations required by the different implementations considered in this work. Here, $p$ is the number of QAOA layers, $N_q$ is the number of data qubits in the 1D chain, and $d$ is the code distance for the QEC-code-inspired Hamiltonians. The qLDPC instances BB18$\,{=}\,$BB$[[18,4,3]]$, BB30$\,{=}\,$BB$[[30,4,5]]$, and BB48$\,{=}\,$BB$[[48,4,7]]$ are taken from \cite{tham2026breakevendemonstrationquantumlowdensity}.}
    \label{Tab:gate-counts}
    \begin{tabular}{|c|c|c|c|}
        \hline
        Prob & $N_{\mathrm{2q}}$ & $N_{\mathrm{mcm}}$ \\
        \hline
        1D 
        & $2p(N_q - 1)$
        & $p(N_q - 1)$ \\
        \hline
        SC
        & $p\left(4d(d - 1)\right)$
        & $p\left(d^2 - 1\right)$ \\
        \hline
        CC
        & $p(d-1)\left(\frac{9(d-3)}{4} + 6\right)$
        & $p(d-1)\left(\frac{3(d-3)}{8} + \frac{3}{2}\right)$ \\
        \hline
        BB18
        & $70p$
        & $14p$ \\
        \hline

        BB30
        & $130p$
        & $26p$ \\
        \hline

        BB48
        & $220p$
        & $44p$ \\
        \hline
    \end{tabular}
\end{table}

The QPUs tested in this work are \texttt{quantinuum\_helios\_1} and \texttt{quantinuum\_h2\_1} from Quantinuum; \texttt{ibm\_fez},\texttt{ibm\_marrakesh},\texttt{ibm\_boston}, \texttt{ibm\_kingston}, \texttt{ibm\_pittsburgh}, and \texttt{ibm\_phoenix}; and \texttt{iqm\_garnet} and \texttt{iqm\_emerald} from IQM. To the best of our knowledge, these are the only publicly available QPUs that allow some degree of MCM implementation on their hardware.

For the QEC-structure evaluation, \texttt{quantinuum\_helios-1}, \texttt{quantinuum\_h2-1}, and \texttt{ibm\_phoenix} are the only QPUs that provided all the elements required to implement the proposed tests directly at the time of our evaluation. Fig.~\ref{Fig:experimental-setup}(c) shows the IBM Nighthawk chip, and different position where the $d=3$ SC is executed. For the IBM QPUs based on the heavy-hex architecture, the limited connectivity prevents a direct mapping of the code structures considered in this work without introducing additional routing operations. On IQM devices, the required dynamic feed-forward operations are restricted by the conditional-control architecture, feed-forward is confined to predefined groups of qubits, and each target qubit can receive conditional control from only one source qubit within a circuit. These constraints prevent the direct implementation of the conditional-control patterns required by the QEC structures considered here.

\subsection{QEC memory experiments setup}
\begin{figure*}[!tbh]
\centering
\includegraphics[width=1\linewidth]{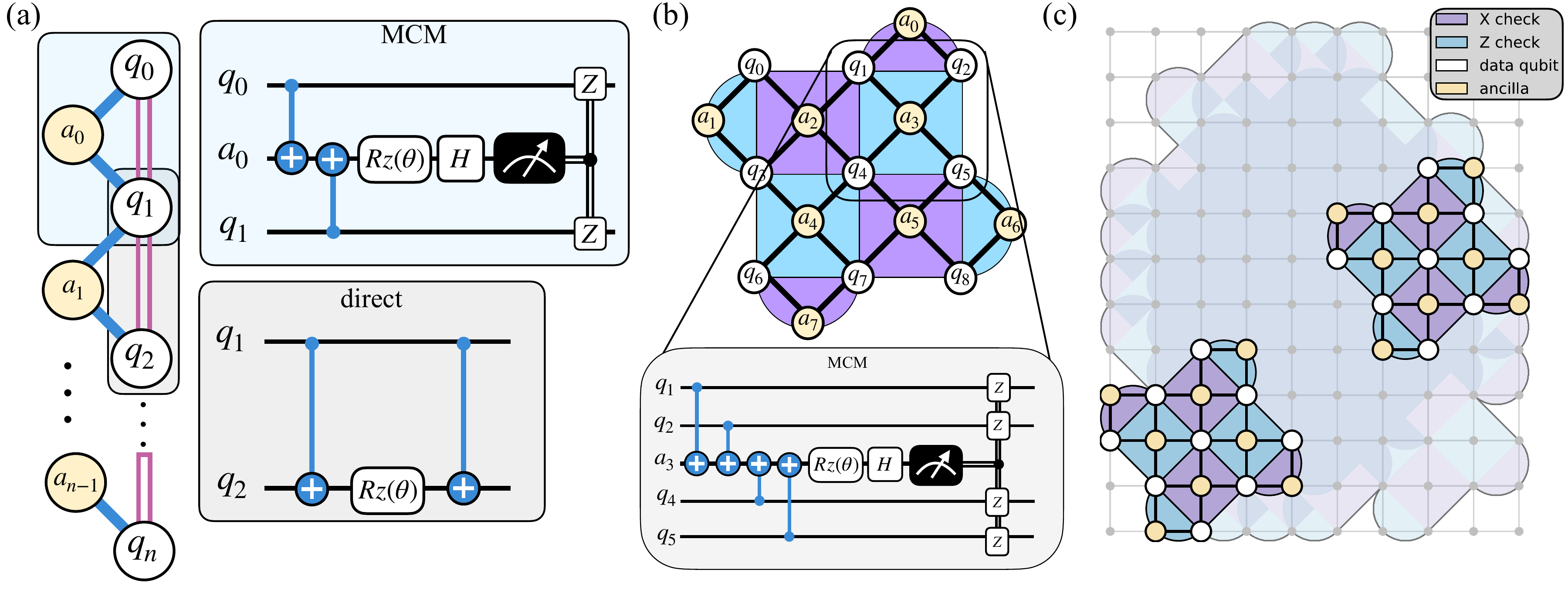}
\caption{\label{Fig:experimental-setup} Circuit representation of the direct and MCM-mediated implementations of the cost Hamiltonian for (a) the 1D chain and (b) the $d=3$ surface code (SC). The LR-QAOA circuit shows the MCM implementation of one representative Hamiltonian term. (c) Distribution of the $d=3$ experiments on \texttt{ibm\_phoenix}, comparing memory and LR-QAOA experiments in both the $Z$ and $X$ bases.}
\end{figure*}

As a final experiment, we implement SC memories on \texttt{ibm\_phoenix} at distances $d=3$, followed by LR-QAOA experiments using the corresponding SC-like structures within the same experimental time window in different parts of the QPU. Fig.~\ref{Fig:experimental-setup}(c) shows the different position where the test is implemented, in total we use 11 different locations, for both the LR-QAOA and memory experiments. The purpose of these experiments is to test whether the performance measured by the LR-QAOA benchmark is correlated with the performance of an actual QEC protocol under similar hardware conditions. For the memory experiment, the data qubits are initialized in the logical state $|0\rangle_L$ or $|+\rangle_L$, followed by $R$ rounds of syndrome extraction and a final transversal measurement of the data register in the $Z$ or $X$ basis respectively. The two runs share the same patch, the same ancillas and the same syndrome-extraction circuit, and differ only in the preparation and the readout
basis, the $|0\rangle_L$ memory is limited by errors that flip the data qubits, which the $Z$ checks detect, and the $|+\rangle_L$ memory by phase errors, which the $X$ checks detect. 

Detectors are constructed following the convention used by \texttt{stim}~\cite{Gidney_2021}. In the
first round, detection events are defined from the stabilizer checks that are deterministic after initialization, i.e., those of the prepared basis. In subsequent rounds, each detector compares the outcome of every stabilizer measurement with its value in the preceding round. After the final round, the checks of the prepared basis are reconstructed from the data-qubit measurements and compared with their last syndrome measurement. Detection events are decoded using minimum-weight perfect matching (MWPM)~\cite{higgott2021pymatching}, with the matching graph obtained from a detector error model generated by \texttt{stim} for the same circuit under uniform circuit-level noise; this strength enters only as the edge weights of the decoder and is not a model of the device. 

A shot is classified as a logical failure when the decoded logical observable disagrees with the measured logical value, and the logical failure probability $\pL$ is the fraction of failed shots over $4000$ repetitions for each patch, basis, code distance and round count $R\in\{2,3,5,7,10\}$. The LR-QAOA experiments on the same patches use $1000$ samples per point at the matching depths $p = R$, and are run in both bases, the $Z$ basis of Sec.~\ref{sec:methods-basis}, giving $r_{\mathrm{ovl}}^{Z}$, and the conjugate $X$ basis, giving $r_{\mathrm{ovl}}^{X}$, where
$r_{\mathrm{ovl}} = (r - r_{\mathrm{rand}})/(r_{\mathrm{ideal}} - r_{\mathrm{rand}})$ in both cases (the two share the same noiseless reference $r_{\mathrm{ideal}}$). The superscript denotes the basis
in which the benchmark is executed, whereas on $\pL^{Z}$ and $\pL^{X}$ it denotes the logical state the memory holds. The two bases are submitted in the same jobs, one immediately after the other, so they see the same device state.

All four measurements run on the same physical data qubits, ancillas and couplers of each of the $11$ at $d = 3$ within a single session. Comparing $r_{\mathrm{ovl}}^{Z}$ and $r_{\mathrm{ovl}}^{X}$ with $\pL^{Z}$ and $\pL^{X}$ across placements, code distances and circuit depths then asks not only whether the structural benchmark follows logical-memory performance, but whether the basis in which the benchmark is executed follows the error channel that limits each memory.

\subsection{Benchmarking protocol}

Here, we provide a guide for how we envision this benchmarking protocol being used. First, individual QPUs can be characterized using the 2-qubit 1D-Chain experiment. For superconducting technologies, it is desirable to run this experiment across the entire chip to assess the homogeneity of MCM operations and identify spatial variations in their performance. Once the MCM operation has been characterized at the level of individual triplets, the 1D-chain test can extended to study how the performance degrades as increasingly larger numbers of qubits and MCM operations are included in the computation.

Finally, specific QEC structures can be benchmarked depending on the target architecture. For instance, superconducting platforms targeting a particular QEC code, e.g. \cite{Bravyi_2024,steffan2025tilecodeshighefficiencyquantum}, can use the corresponding structure to evaluate how successive generations of their devices improve in executing the operations required by that code using both X and Z bases. Similarly, trapped-ion architectures moving toward modular or limited-connectivity, e.g. \cite{tripier2026faulttolerantquantumcomputingtrapped}, designs can be evaluated using structures representative of their target architecture. In this way, the benchmark provides different levels of characterization, from local MCM performance to large-scale QEC structures, so improvements across hardware generations can be tracked and, ultimately, determined whether these physical improvements translate into improved logical performance.

\section{Results}\label{Sec:Results}

\subsection{The minimum MCM experiment}\label{Sec:1d-2q}

\begin{figure*}[!tbh]
\centering
\includegraphics[width=1\linewidth]{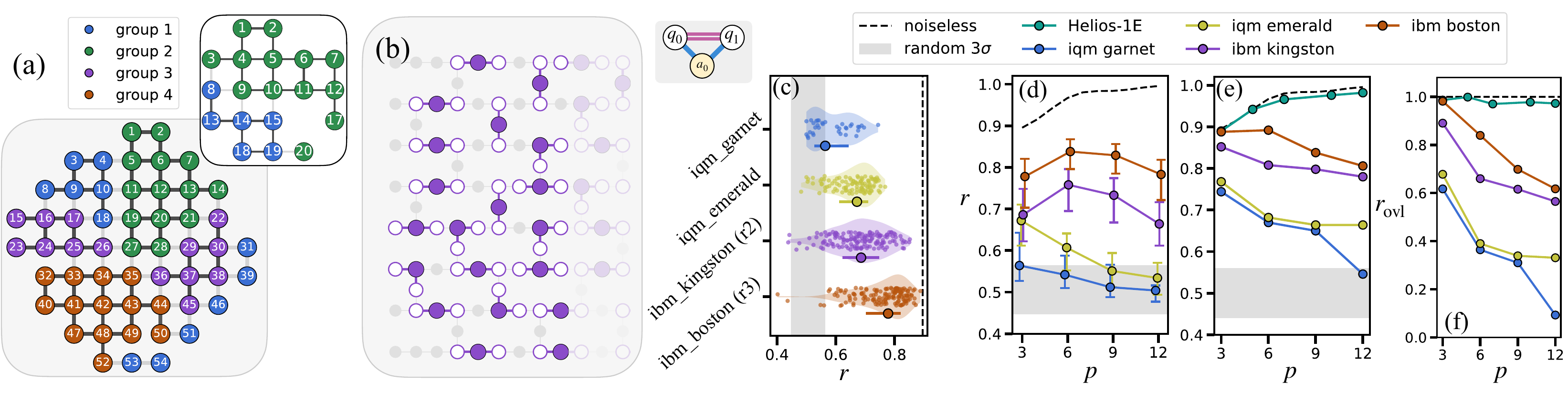}
\caption{\label{Fig:1D-2q} 2-qubit MCM experiment on different QPUs. (a) Layouts of \texttt{iqm\_garnet} (upper right) and \texttt{iqm\_emerald} (lower left), with different colors indicating pairs of qubits between which a MCM can be applied. (b) IBM Heron heavy-hex layout. Highlighted qubits indicate those used in one of the triplet experiments. The filled circle denotes the ancilla qubit on which the MCM is applied, while the connected qubits are the other qubits involved in the experiment. (c) $r$ distribution of the different triplets used on each QPU for the MCM implementation of LR-QAOA at $p=3$. (d) average results at different circuit depths for the same QPUs. Error bars represent one standard deviation. (e) Best triplet evolution on each QPU, additional results are presented here for the noisy emulator \texttt{quantinuum\_helios-1E}. (f) $r_\mathrm{ovl}$ for the triplets in panel (e).}
\end{figure*}

The 1D experiments provide a primitive-level control for the MCM-mediated LR-QAOA construction used in the main text. Figure~\ref{Fig:1D-2q} shows the minimal two-data-qubit experiment used to evaluate the MCM operations and provides a diagnostic of their performance across the full QPU with a reduced number of experiments. Figure~\ref{Fig:1D-2q}(a) shows the layouts of the IQM QPUs. On these QPUs, there is a restriction on how qubits can be combined to perform an MCM operation, two qubits belonging to different groups cannot share the same MCM operation. 

Figure~\ref{Fig:1D-2q}(b) shows the circuit layout used on IBM's heavy-hex devices. The highlighted triplets are executed in parallel. This procedure is repeated three times with different ancilla qubits, such that all qubits that can act as ancillas are tested. A similar protocol is used on IQM devices, maximizing the number of triplets executed in parallel in each run.

Figure~\ref{Fig:1D-2q}(c) shows the MCM quality across every ancilla-capable qubit of four QPUs. $H = Z_{d_0}Z_{d_1}$, evaluated independently for every ($d_0$, $a_0$, $d_1$) triple in which $a_0$ is coupled to both data qubits. Each dot is one triple, 148 triples on each 156-qubit Heron device, 104 on \texttt{iqm\_emerald}, and 35 on \texttt{iqm\_garnet}. The grey band is the $3\sigma$ interval of uniform random sampling ($r_\mathrm{rand} = 0.506 \pm 0.020$) and the dashed line the exact noiseless value $r_\mathrm{ideal} = 0.895$. $\Delta_{\gamma,\beta} = 0.5$, 500 shots per triple. The violin plots show the distribution at p=3. A summary of the main results and additional QPUs not shown in the figures are given in Table~\ref{tab:2q-summary}, including the median $r$ and interquartile range (IQR), as well as the fraction of triplets above the random threshold ($>3\sigma$), and the $r_\mathrm{ovl}$ at depth 3 and 12.

Fig.~\ref{Fig:1D-2q}(d) shows the median and IQR at different depths. Here, the separation between \texttt{iqm\_garnet} and \texttt{iqm\_emerald} is important, indicating that a larger fraction of qubits in the newer generation performs better than in the previous one. However, their performance still degrades faster with circuit depth than that of the IBM QPUs. At $p=12$, only \texttt{ibm\_boston} and \texttt{ibm\_kingston} retain a signal above the random threshold. Additionally, IBM QPUs performance continues to improve to $p=6$, indicating that, at these depths, the algorithmic signal of LR-QAOA is stronger than the accumulated circuit noise. Comparing successive QPU generations, the transition from Heron revision r2 (\texttt{ibm\_kingston}) to r3 (\texttt{ibm\_boston}) shows a substantial improvement, similar to that observed from \texttt{iqm\_garnet} to \texttt{iqm\_emerald}.

Figures~\ref{Fig:1D-2q}(e)-(f) show the best ancilla on each QPU, to separate hardware limits from chip inhomogeneity. \texttt{quantinuum\_helios-1E} (Helios-1E) is included as a baseline for how this circuit performs on trapped-ion hardware. For the two-qubit case, it maintains $r_\mathrm{ovl}$ between 0.97 and 1.00 across $p=3$-$12$ ($r=0.890$-$0.982$). The separation between the IBM QPU revisions persists for the best-performing triplet, indicating an improvement in the newer device. By comparison, the best triplets on the two IQM QPUs follow similar trends.

\begin{table}[t]
\centering
\caption{Summary of the two-qubit LR-QAOA experiments. $r$ and $>3\sigma$ are taken at $p=3$ over all data-ancilla-data triples; $r_{\rm ovl}^{3\rightarrow12}$ is the mean
normalized overlap over the same triples at $p=3$ and $p=12$.}
\label{tab:2q-summary}
\scriptsize
\setlength{\tabcolsep}{3pt}
\begin{tabular}{|lccc|}
\hline
QPU & $r$ (IQR) & $>3\sigma$ & $r_{\rm ovl}^{3\rightarrow12}$\\
\hline
\texttt{ibm\_boston}     & 0.778 (0.703--0.821) & 97\% & 0.65$\rightarrow$0.52\\
\texttt{ibm\_pittsburgh} & 0.755 (0.660--0.808) & 90\% & 0.57$\rightarrow$0.46\\
\texttt{ibm\_kingston}   & 0.686 (0.622--0.748) & 89\% & 0.46$\rightarrow$0.32\\
\texttt{ibm\_phoenix}    & 0.558 (0.471--0.653) & 48\% & 0.17$\rightarrow$0.23\\
\texttt{ibm\_fez}        & 0.715 (0.639--0.754) & 87\% & 0.47$\rightarrow$0.26\\
\texttt{ibm\_marrakesh}  & 0.665 (0.572--0.724) & 78\% & 0.37$\rightarrow$0.21\\
\texttt{iqm\_emerald}    & 0.672 (0.612--0.710) & 86\% & 0.40$\rightarrow$0.06\\
\texttt{iqm\_garnet}     & 0.564 (0.527--0.643) & 49\% & 0.22$\rightarrow$0.00\\
\hline
\end{tabular}
\end{table}

\subsection{MCM as a perceived contribution of noise}\label{Sec:res-mcm}

We first quantify how expensive the MCM primitive is on today's devices increasing 1D-Chains of qubits. LR-QAOA has a characteristic noise structure (Methods, Sec.~\ref{Sec:noise}) that can be map to a depolarizing noise onto a single exponential Eq~.\ref{eq:hardware_model} structure, so a hardware run can be assigned the depolarizing rate $\lambdaeff$ that reproduces a similar degradation. Because a direct and an MCM implementation has the same circuit logic and differ only by the inserted MCMs, the gap between them shows the impact of using MCM operations and feedforward their information. We characterize this by a perceived $\lambda_{\mathrm{mcm}}$.

Results of IQM QPUs are not included because of a second restriction that appears when considering a three-data-qubit chain or more. This require a layout of the form $q_0$--$a_0$--$q_1$--$a_1$--$q_2$. In this configuration, the central data qubit, $q_1$, must couple to two different ancillary qubits that perform MCMs and provide the feedforward information required for the logical controlled-$Z$ rotation at the end of the operation [Fig.~\ref{Fig:experimental-setup}]. Because of this connectivity restriction, 1D chains containing more than two data qubits cannot be implemented. This connectivity constraint is not, per se, a fundamental limitation for implementing a logical-memory QEC experiment. It does, however, restrict the implementation of adaptive fault-tolerant protocols that require measurement-conditioned operations. For example, magic-state cultivation requires feedback during state preparation, while the subsequent implementation of a logical T gate through magic-state injection requires a real-time, measurement-dependent Clifford correction.

\begin{figure*}[!tbh]
\centering
\includegraphics[width=1\linewidth]{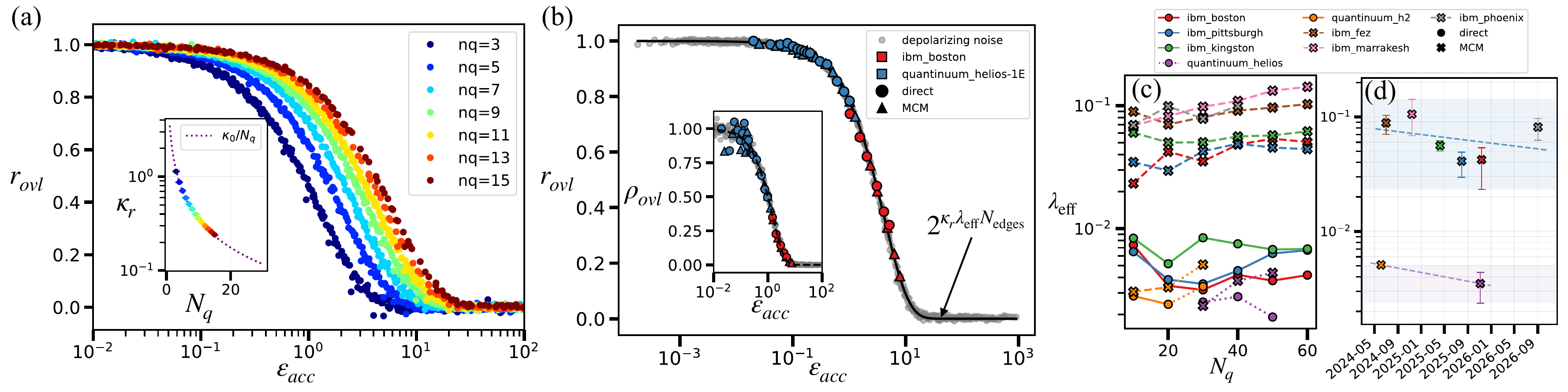}
\caption{\label{Fig:1D_lambda} 1D-Chain experiments used to construct an equivalent depolarizing-noise model. (a) $r_\mathrm{ovl}$ as a function of depolarizing-noise strength, LR-QAOA depth, and problem size. The inset shows the fitted $\kappa$ for each problem size; the dotted curve is the scaling $\kappa_0/N_q$. (b) $r_\mathrm{ovl}$ versus accumulated error for the $10$-qubit problem. Gray background points show the depolarizing-noise simulations. Hardware data from \texttt{ibm\_boston} and \texttt{quantinuum\_helios-1E} are shown for $p\in\{10,15,20,30,40,50\}$, for both direct and MCM implementations. The solid black curve corresponds to the fitted approximation-ratio susceptibility $\kappa_r=3.69/N_q$. The inset shows the success-probability overlap for the same conditions, with fitted $\kappa_\rho=1.0$. (c) $\lambda_{\mathrm{eff}}$ versus number of qubits for different QPUs, comparing direct and MCM implementations. (d) Average effective rates for mid-circuit measurement experiments as a function of QPU availability date.
Error bars indicate the standard deviation. For  \texttt{quantinuum\_h2-1}, the average includes only experimental data, i.e., points for $N_q>20$.}
\end{figure*}

Figure~\ref{Fig:1D_lambda} summarizes the one-dimensional analysis used to turn LR-QAOA decay into an equivalent depolarizing-noise model. Fig.~\ref{Fig:1D_lambda}(a) shows the calibration procedure using different depolarizing noise strengths and chain sizes $N_q=3$-$15$. Fitting the different curves of $r_\mathrm{ovl}$, are well characterized by Eq.\ref{eq:hardware_model}, but we notice a shift of their $\kappa_r$. The inset of Fig.~\ref{Fig:1D_lambda}(a) shows that this shift is inversely proportional to the number of qubits, with $\kappa_r=(3.69\pm0.04)/N_q$. This inverse-size scaling indicates that the approximation-ratio signal becomes less sensitive to accumulated error as the cost function becomes more extensive. One can also obtain a similar estimate for the success probability, although the corresponding $\rho_\mathrm{ovl}$ is not shown. Its susceptibility saturates near $\kappa_\rho\simeq0.965$.

Figure~\ref{Fig:1D_lambda}(b) shows a 10-qubit experiment on \texttt{ibm\_boston} and simulations on \texttt{quantinuum\_helios-1E}, using both direct and MCM-mediated implementations. Using the depolarizing-noise model, we can assign an effective error rate that produces the same signal decay as the observed data. This does not mean that the hardware noise is purely depolarizing; but, the model provides a single effective parameter with which to quantify the observed decay, using a noise process that is well characterized for both QPUs. The inset of Fig.~\ref{Fig:1D_lambda}(b) shows the complementary $\rho_\mathrm{ovl}$. For the $N_q=10$ \texttt{ibm\_boston} data, $r_\mathrm{ovl}$ and $\rho_\mathrm{ovl}$ infer consistent estimates of $\lambda_\mathrm{eff}$: $\lambda_\mathrm{direct(mcm)}=0.01078,(0.01656)$ and $0.01080,(0.01680)$, respectively, corresponding to differences of only $0.2\%$ and $1.5\%$. This agreement is nontrivial, while both observables follow the depolarizing-noise model in simulation, there is no requirement that realistic hardware noise produce the same effective depolarizing rate when inferred from two different observables. Their consistency therefore might indicate that the leading hardware-induced degradation is well captured by a single depolarizing-equivalent parameter across both the approximation ratio and the success probability.

Figure~\ref{Fig:1D_lambda}(c) extends the results of (b) to problem sizes from 10 to 60 qubits. We use 1000 and 50 shots on IBM and Quantinuum devices, respectively. The direct 1D-chain implementation gives an approximately size-independent $\lambda_\mathrm{eff}$, about $0.2\%$ ($0.17$-$0.28\%$) on the \texttt{quantinuum\_helios-1} data and $0.30$-$0.80\%$ across the IBM devices, where the smallest chain ($N_q = 10$) sits at the upper end of that range. The MCM contribution is more important. On IBM devices the perceived MCM cost is already large at small size and varies only weakly with problem size, over $N_q = 10$-$60$ it changes by $1.2\times$ on \texttt{ibm\_kingston} and $1.5\times$ on \texttt{ibm\_fez}, and by at most $2.3\times$ on
\texttt{ibm\_boston}. In contrast, the MCM contribution for \texttt{quantinuum\_helios-1} grows with chain length, from $\lambda_\mathrm{eff} = 0.22 \pm 0.04\%$ at $N_q = 30$ to $0.41 \pm 0.05\%$ at $N_q = 50$, a factor of $1.9$. That longer chains raise $\lambda_\mathrm{eff}$ is what one expects if the layer is partially serialized \texttt{quantinuum\_helios-1} has eight
operation zones and processes at most sixteen qubits simultaneously, so the $N_q-1$ mid-circuit
measurements of a layer cannot all be performed at once, and the data qubits not being measured
idle while the rest of the layer completes.

At $N_q = 30$, we can directly compare \texttt{quantinuum\_h2-1} and \texttt{quantinuum\_helios-1}. At this problem size, we obtain $\lambda_\mathrm{eff} = 0.49 \pm 0.24\%$ for \texttt{quantinuum\_h2-1}, compared with $0.22 \pm 0.04\%$ for \texttt{quantinuum\_helios-1}. The central values therefore indicate a reduction in the effective error, although the uncertainties are large for this measurement alone to conclusively resolve an improvement between the two hardware generations. Nevertheless, the trend is consistent with the subsequent experiments presented in the following sections.

Figure~\ref{Fig:1D_lambda}(d) summarizes the results shown in panel (c), with the error bars representing the mean and standard deviation across the qubit chains. The devices are ordered according to the date on which they became available through the cloud. For the IBM QPUs, and consistently with the trends observed in the triplet experiments, newer device generations generally exhibit lower effective error rates up to \texttt{ibm\_boston}. This trend changes for the subsequent generation, which introduces the square-lattice topology of the Nighthawk architecture. The corresponding change in connectivity and circuit mapping may contribute to the larger $\lambda_\mathrm{eff}$ observed for the latest device, making a direct comparison with the preceding generations less straightforward. For Quantinuum, \texttt{quantinuum\_helios-1} exhibits a lower effective error rate than \texttt{quantinuum\_h2-1}, suggesting that the newer generation also provides improved performance for the MCM-intensive circuits considered here.

\subsection{Quantifying performance changes over time}

\begin{figure}[!tbh]
\centering
\includegraphics[width=0.8\linewidth]{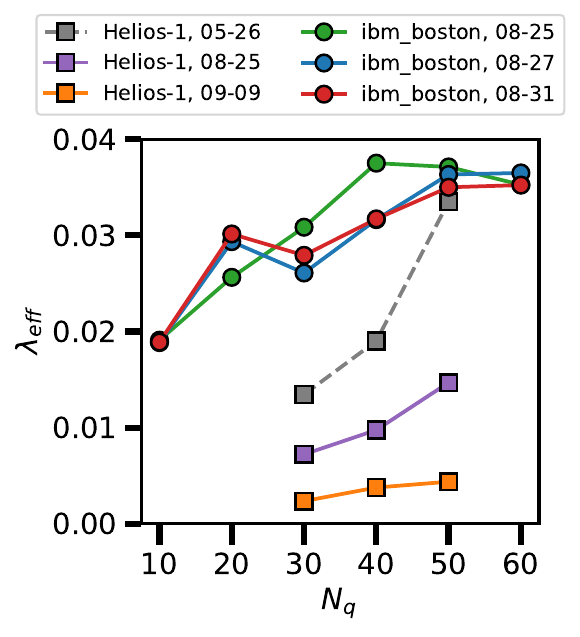}
\caption{\label{Fig:time} 1D-chain MCM experiments performed at different times on IBM and Quantinuum QPUs. The labels indicate the month and day of 2026 on which each experiment was submitted. For \texttt{ibm\_boston}, the data shown in green correspond to a different subset of qubits, and the other two datasets use the same qubit chain. Experiments on \texttt{ibm\_boston} were performed with 1000 shots, while those on \texttt{quantinuum\_helios-1} used 50 shots.
}
\end{figure}

Figure~\ref{Fig:time} shows the MCM 1D-chain experiment repeated at different points in time. For \texttt{quantinuum\_helios-1}, the effective error rate decreases relative to the first set of experiments. The first shift (gray to purple) may reflect improvements in calibration or other system-level optimizations introduced between the two runs, whereas the second shift (orange curve) is attributable to improvements in our implementation. The first two experiments on Helios-1 used a serialized implementation that did not exploit the eight available zones in which two-qubit gates and MCMs can be executed concurrently. As a result, both two-qubit gates and MCM operations were performed largely sequentially. For the orange experiment, we modified the implementation to parallelize two-qubit gates and MCM operations whenever possible. The resulting performance improvement is clearly visible in the curves with only 50 shots. These results highlight two complementary capabilities of the benchmark, it is sensitive both to improvements in the underlying system, such as calibration changes, and to improvements in scheduling and implementation efficiency.

For \texttt{ibm\_boston}, the results remain consistent across different experimental runs. This temporal stability is desirable, as it suggests that the extracted effective error rate provides a representative measure of the average QPU performance rather than being dominated by short-term fluctuations. Moreover, the agreement between runs performed on different subsets of qubits suggests a comparable level of performance across different regions of the device. As the chain size increases, the subsets increasingly share the same physical qubits, which also explains the convergence of the corresponding data points for the largest chains.

\subsection{QEC structure evaluation}\label{Sec:res-structure}
We use LR-QAOA as a preliminary, low-cost test of how a device executes a QEC code's own structure. For each family, the cost Hamiltonian is built from the code's checks (Eq.~\eqref{Eq:CC} for the colour code), and its MCM implementation reproduces the circuit of a syndrome-extraction round. Each LR-QAOA layer $p$ therefore plays the role of one syndrome-extraction round, and sweeping $p$ probes how the code's connectivity and measurement pattern accumulate error on a specific device. The approximation ratio $r=0.5$ is the random-sampling limit of every check Hamiltonian considered here; a QPU reaching it has accumulated enough error to erase all algorithmic signal.

\begin{figure}[!tbh]
\centering
\includegraphics[width=1\linewidth]{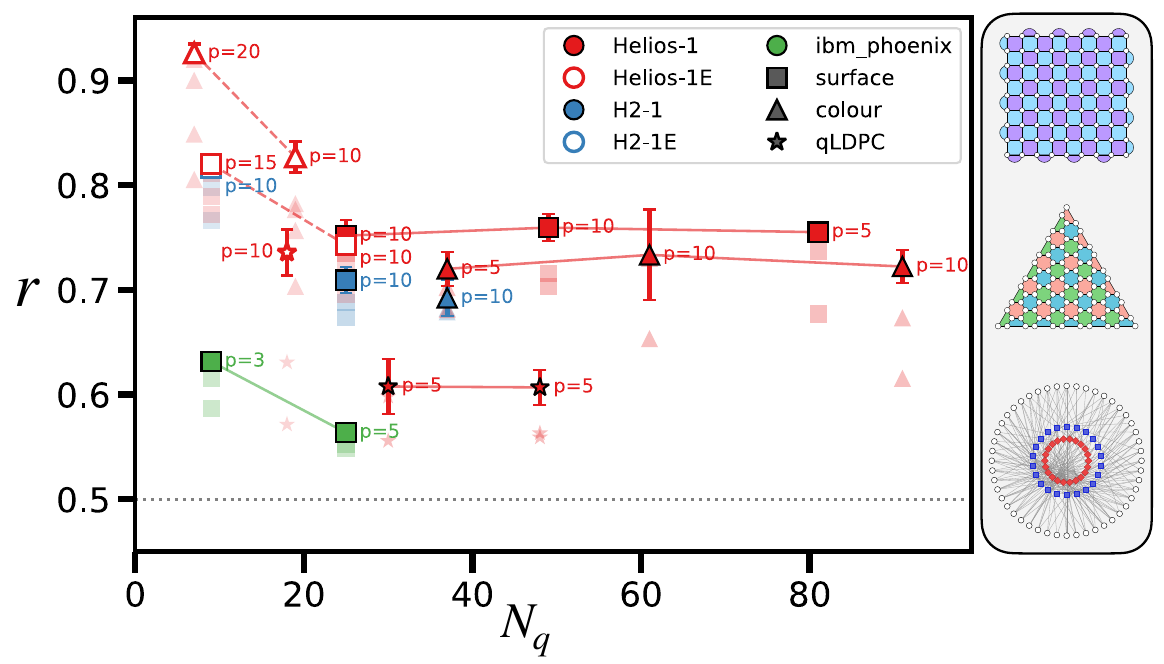}
\caption{\label{Fig:QEC_exp}QEC structure evaluation. Approximation ratio $r$ against the number of data qubits $N_q$ for the MCM implementation of the check Hamiltonians of three code families: triangular colour code $d=3$-$11$ ($7$-$91$ data qubits), surface code $d=3$-$9$ ($9$-$81$ data qubits) and bivariate-bicycle qLDPC codes BB$[[18,4,3]]$, BB$[[30,4,5]]$ and BB$[[48,4,7]]$. Marker shape is the code family and colour is the machine; emulators are hollow. Each highlighted point is the best depth measured for that structure; faint points are the other depths measured at the same size. The dotted line at $r=0.5$ is the random-sampling limit, identical for every code here. Error bars are one standard deviation of shot noise ($10$-$50$ shots on the trapped-ion runs, $1000$ on \texttt{ibm\_phoenix}). Annotations refer to the depth of the best $r$. The structures of the largest member of each family are drawn in the right-hand panel, with circles and colors indicating data qubits and the corresponding syndrome-extraction checks, respectively.}
\end{figure}

Figure~\ref{Fig:QEC_exp} shows this evaluation for three code families, 3 QPUs, and 2 emulators. On the trapped-ion machines the colour code stays well above the random limit from $d=3$ to $d=11$ ($7$ to $91$ data qubits), reaching $r=0.926$ at $d=3$ and $r=0.733$ at $d=11$ ($8.1\sigma$ above random with $10$ shots). At $d=11$, the $91$ data qubits leave only $7$ of the machine's $98$ physical qubits available as ancillas, so the $45$ ancilla measurements required per layer must be distributed over nine batches of at most seven checks, rather than the three disjoint measurement classes allowed by the code structure. The surface code behaves likewise, $r=0.752$ at $d=5$ and $r=0.737$ at $d=9$ ($19.0\sigma$ with $20$ shots), so $r$ degrades only weakly between $25$ and $81$ data qubits. At $d=9$, SC contains $80$ checks per layer but leaves $17$ physical qubits available as ancillas; its execution requires seven batches and nevertheless retains a strong LR-QAOA signal.

The bivariate-bicycle qLDPC codes probe a different structural regime. Every check has weight $5$ and wraps around the $\ell\times m$ torus, so connectivity and measurement pattern are intrinsically non-local. Here ancilla space plays no role, BB48 leaves $50$ qubits free, yet the checks overlap so heavily that a layer still needs seven batches of at most eight checks, a serialisation set by the code, not by the machine. All three structures run above the random limit on \texttt{quantinuum\_helios}, $r=0.736$ for BB18, $0.608$ for BB30 and $0.607$ for BB48 ($4.9\sigma$ and $6.3\sigma$ at $20$ shots), showing that circuits built on these qLDPC structures still yield an algorithmic signal, and providing a basis for comparison with the local codes.

Two additional diagnostics emerge from these experiments. First, the depth at which $r$ reaches its maximum provides a measure of how far the LR-QAOA signal can build up before accumulated noise dominates. On the emulators, $r$ continues to improve up to $p=15$-$20$, whereas on hardware the optimum typically occurs around $p=5$-$10$ and, on \texttt{ibm\_phoenix}, already at $p=3$--$5$. Second, superconducting and trapped-ion platforms can be directly compared at fixed code structure. For the surface code at $d=3$, we obtain $r=0.820$ on \texttt{quantinuum\_helios\_1e} and $r=0.631$ on \texttt{ibm\_phoenix}, while at $d=5$ the corresponding values are $0.752$ and $0.564$, respectively, with both platforms remaining significantly above the random-sampling limit. This structural information is obtained using only $10$-$50$ shots per point on the Quantinuum devices. 

This test provides a useful measure of progress over time as more codes and QPU-representative instances are added. Improvements should appear both as an increase in the number of qubits that can be tested and in the maximum $r$ achieved at a larger depth, so we expect the plot to progressively extend along both the $x$ and $y$ directions.

\subsection{Sampling budget to draw conclusions}\label{Sec:sampling-budget}
\begin{figure}[!tbh]
\centering
\includegraphics[width=1\linewidth]{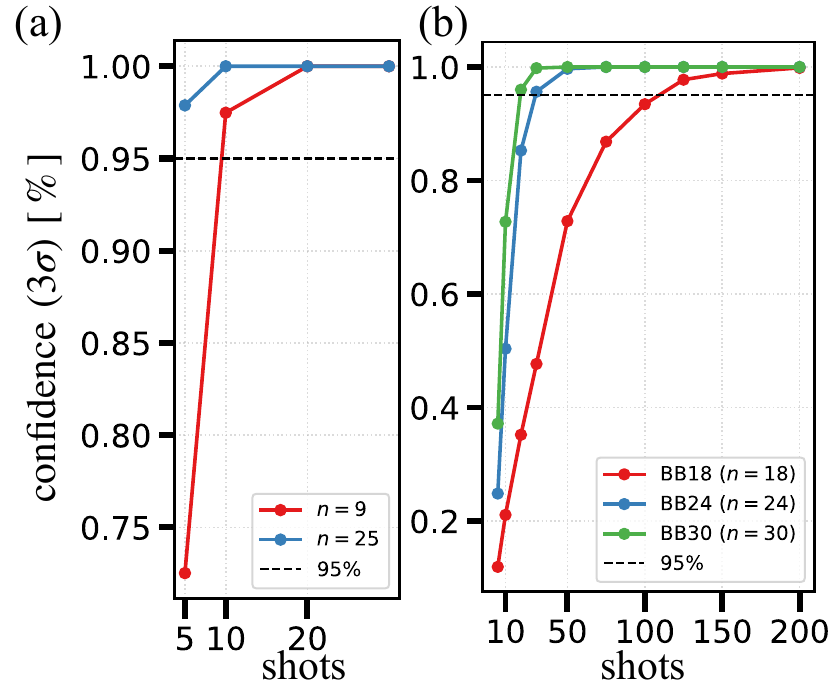}
\caption{\label{Fig:random_budget} Sampling budget to distinguish the given LR-QAOA samples from a random sampler with a $3\sigma$ confidence for (a) SC structure (b) the qLDPC families.
}
\end{figure}

Figure~\ref{Fig:random_budget} quantifies the number of samples required to reject the random-sampling hypothesis for a $p=3$ LR-QAOA implementation of
surface-code and BB-qLDPC Hamiltonians. This is a useful diagnostic because it tells if the noisy device still preserves algorithmic information, or has the output already collapsed to the random limit.

For the SC instances, only a few to tens of shots are sufficient to distinguish the noiseless LR-QAOA distribution from random sampling at the $3\sigma$ level. For the BB-qLDPC family the required budget is larger for the smallest instance, BB18, but decreases for the larger BB24 and BB30 instances. Thus, within the simulated families, increasing the code size does not make the random-baseline test harder; in fact, the larger instances are often easier to certify because their $p=3$ LR-QAOA signal is more clearly separated from the random sampler and random signal variance shrinks as problem size grows (See Sec.~\ref{Sec:res-rank}).

This observation is important for interpreting the hardware data. In Fig.~\ref{Fig:QEC_exp}(b), when a surface-code LR-QAOA point approaches the random limit, the random-budget analysis shows that this behavior cannot be explained simply by insufficient sampling. At these sizes, a non-random $p=3$ signal would already be visible with the available shot budget. Therefore, failure to separate from the random baseline is itself a meaningful indication
that the implemented circuit has reached the noise-dominated regime.  Conversely, when the measured approximation ratio remains above the random threshold, even with few samples, it provides statistically supported evidence that the device
still preserves code-structured LR-QAOA information.

To estimate the sampling budget required to distinguish two depolarizing-noise models, following Sec.~\ref{Sec:res-rank}, we use simulations of the SC $d=3$ LR-QAOA benchmark with $50{,}000$ shots per device and depth. This makes the inferred distinguishability threshold $S^\ast$ stable to within a few percent. 

\begin{figure}[!tbh]
\centering
\includegraphics[width=1\linewidth]{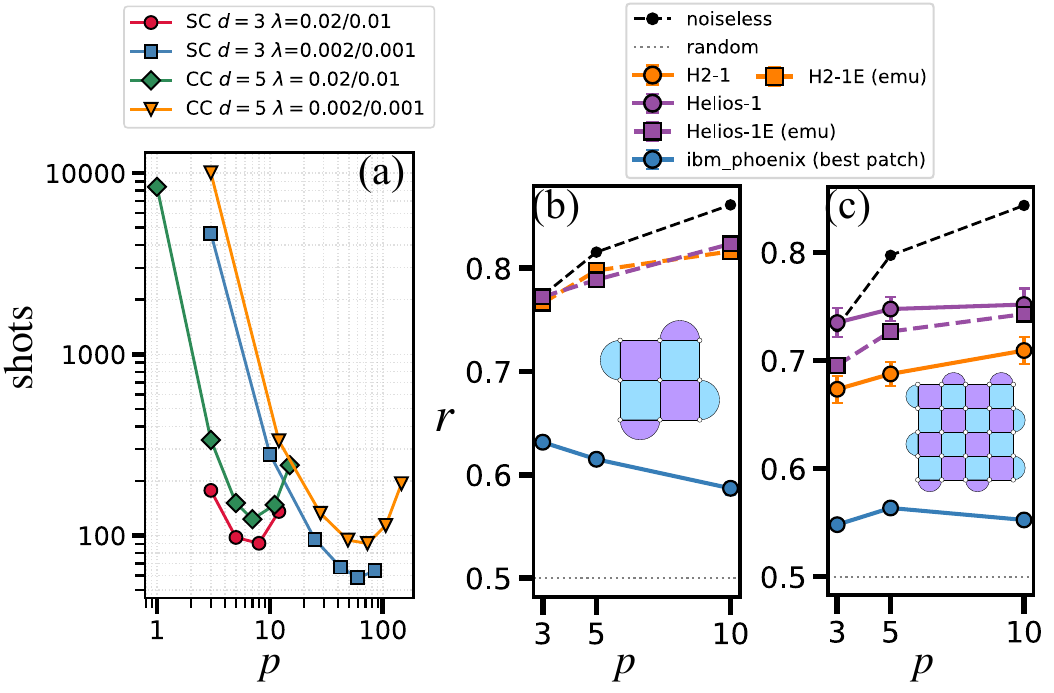}
\caption{\label{Fig:distinguish_budget}Sampling budget to distinguish two QPUs and experimental realization on QPUs. (a) Depolarizing-noise simulations of the samples required to distinguish $\lambda_A$ from $\lambda_B$ at $3\sigma$ confidence, for a color code with $d = 5$ and a surface code with $d = 3$. (b) Approximation ratio $r$ of the surface-code-like structure of distance $d = 3$ (9 data qubits) on the Quantinuum emulators and on \texttt{ibm\_phoenix}, and (c) $d = 5$ (25 data qubits) on \texttt{quantinuum\_h2-1},\texttt{quantinuum\_helios-1} and \texttt{ibm\_phoenix}. Error bars are one standard deviation of shot noise; the dashed black curve is the noiseless value and the dotted line the random-sampling limit $r = 0.5$. Each experimental point uses 50 samples on the Quantinuum QPUs in (c), 500 on the emulators in (b), and 1000 on \texttt{ibm\_phoenix}.
}
\end{figure}

Figure~\ref{Fig:distinguish_budget}(a) shows the sampling budget required to distinguish two QPUs under depolarizing-noise simulations. First, consider QPU 1 with $\lambda_1=0.02$ and QPU 2 with $\lambda_2=0.01$. The two devices can be distinguished at $3\sigma$ confidence using $S^\ast=90$ shots per device for the surface code with $d=3$, and slightly more than $100$ shots for the color code, with the minimum occurring near $p=8$. We then consider two lower-noise QPUs, with $\lambda_1=0.002$ and $\lambda_2=0.001$. In this case, the minimum sampling budget is reduced to $S^\ast=58$ shots, at the cost of running substantially deeper circuits, with the minimum occurring near $p=60$. This behavior is consistent with the predicted scaling $p^\ast\propto1/\lambda$: as the noise level decreases, deeper circuits are required to reach the same transition region in the overlap decay where the two devices become most distinguishable. 

Figures~\ref{Fig:distinguish_budget}(b) and (c) show the same surface-code structure executed on hardware and on the Quantinuum emulators. Here we plot the approximation ratio $r$, together with the noiseless value and the random-sampling limit $r = 0.5$. At $d = 3$ [Fig.~\ref{Fig:distinguish_budget}(b), 9 data qubits, 500 samples on the emulators and 1000 on \texttt{ibm\_phoenix}], both \texttt{quantinuum\_h2-1e} and \texttt{quantinuum\_helios-1e} follow the noiseless curve closely, reaching $r = 0.816$ and $0.824$ at $p = 10$ against the noiseless $0.861$. On \texttt{ibm\_phoenix} the larger error budget of the MCM operations leads to a substantially lower $r$, which decreases with depth, from $0.631$ at $p = 3$ to $0.587$ at $p = 10$, instead of increasing as the noiseless curve does. The signal nevertheless remains $16$-$24$ standard deviations above the random-sampling limit, so algorithmic information is still present despite the stronger MCM-induced degradation.

At $d = 5$ [Fig.~\ref{Fig:distinguish_budget}(c), 25 data qubits, 50 samples on the Quantinuum QPUs and 1000 on \texttt{ibm\_phoenix}], \texttt{quantinuum\_helios-1} lies above \texttt{quantinuum\_h2-1} at every depth, $r = 0.735$, $0.748$ and $0.752$ against $0.673$, $0.688$ and $0.709$, a separation of $2$-$4$ standard deviations per point obtained with only 50 samples. This difference could come from different hardware improvements, including lower two-qubit gate error rates, a larger number of MCM operations that can be executed in parallel, and other architectural improvements affecting the execution of dynamic circuits. The \texttt{quantinuum\_helios-1e} emulator tracks the hardware within the same uncertainty, slightly below it at $p = 3$-$5$, indicating that its noise model is a faithful, mildly conservative description of the device at this size. \texttt{ibm\_phoenix} again sits lowest, $r = 0.548$-$0.564$, yet still $15$-$20$ standard deviations above random.

\subsection{QEC code experiment vs. LR-QAOA estimation}\label{Sec:qec-vs-qaoa}

Finally, we compare the LR-QAOA response directly with SC memory experiments performed on \texttt{ibm\_phoenix}. We implement $Z$ and $X$ bases SC memories at distances $d=3$ for $R\in\{1,2,3,5,7,10\}$ rounds over $11$ different patches, respectively over 1 week, using $4000$ shots per memory experiment. LR-QAOA experiments on the corresponding SC-like structures are performed within the same experimental time window with $1000$ shots using both the $Z$ and $X$ bases for the Hamiltonian.

\begin{figure*}[!tbh]
\centering
\includegraphics[width=1\linewidth]{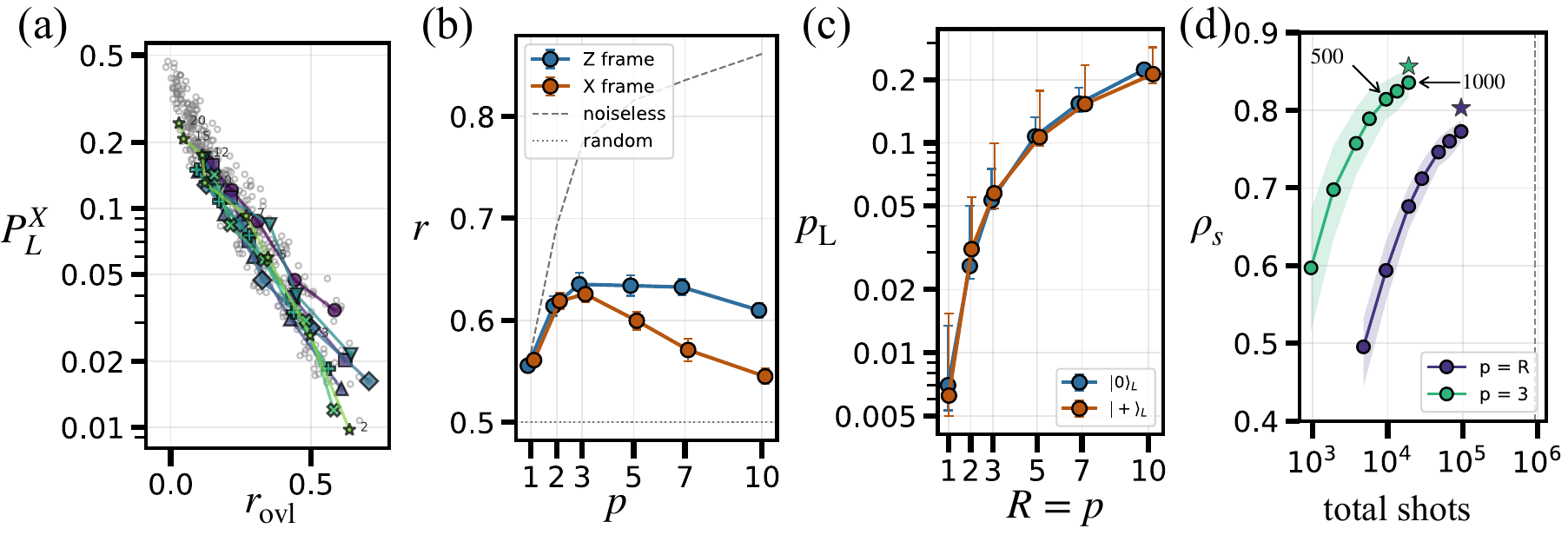}
\caption{\label{Fig:QEC_experiment} Comparison between the logical memory and LR-QAOA benchmark on $d=3$ surface-code patches of \texttt{ibm\_phoenix}, over repeated sessions. (a) Logical error probability after storing $|+\rangle_L$ for $R$ rounds versus the $X$-frame overlap $r_{\rm ovl}$ at matched depth $p=R\geq2$. Colored curves track the patch with the lowest $P_L^X$ in each of experiments; gray circles show all remaining patches. (b) Approximation ratio versus depth for the same patch in the $X$ and $Z$ frames, averaged over sessions with standard-deviation error bars, together with noiseless and random-sampling references. (c) Logical error probability versus rounds for $|0\rangle_L$ and $|+\rangle_L$; each session is summarized by the median over 11 patches, and bars span the 16th--84th percentile across the experiments. (d) Spearman rank correlation between the $X$-frame benchmark and $P_L^X$ versus benchmark shot budget, obtained by resampling shots with replacement; bands show the 10th--90th percentile. Blue uses all matched depths $p=R$, while green uses only $p=3$ against every $R$. Stars show the correlation from all 1000 shots, annotations mark 500 shots per circuit, and the dashed line indicates the shot cost of the memory experiment.}
\end{figure*}

Figure~\ref{Fig:QEC_experiment}(a) compares the LR-QAOA benchmark in the $X$ frame with the decoded memory experiment on the same patches. Each point corresponds to one patch at matched depth $p=R$, comparing its overlap $r_{\rm ovl}^X$ with the logical error probability $P_L^X$ after storing $|+\rangle_L$ for $R$ rounds. The highlighted trajectories follow the patch with the lowest $P_L^X$ across the eight sessions, while the grey points show all other patches. 

The data follow an approximately exponential relation, $\log_{10}P_L^X \approx a-b,r_{\rm ovl}^X$ with $b\approx2$, consistently across patches, sessions, and depths. This is expected since the undecoded benchmark is sensitive to physical errors, while the $d=3$ code corrects single errors, allowing $r_{\rm ovl}^X$ to vary considerably while $P_L^X$ remains small. This is desired as hardware improves, resolving $P_L$ to fixed relative precision requires a number of shots scaling as $1/P_L$, whereas the shot noise of $r_{\rm ovl}$ does not. Thus, the benchmark could provide a cheaper characterization when memory experiments become costly to resolve, provided the relation observed here for $10^{-2}\lesssim P_L\lesssim0.5$ persists at lower error rates.

Figure~\ref{Fig:QEC_experiment}(b) shows the LR-QAOA approximation ratio versus depth in the $Z$ and $X$ frames for the same patch over one week. Two observations follow. First, the sessions agree closely, the day-to-day spread remains within 6-9\% of $r-r_{\rm rand}$ in the $Z$ frame and within 5\% at shallow depth in the $X$ frame, increasing to 18\% at $p=10$, indicating that both the patch and the benchmark are stable over this timescale. Second, the two frames coincide at $p=1$ but separate systematically with depth, with the $X$ frame decaying faster; the same behavior is observed on the other patches. The $Z$-frame measurements exhibit a stable readout bias toward $|0\rangle$ that might have implication in the observed behavior. The growing gap also cannot be attributed to circuit cost, as both frames compile to identical physical gate counts, or to a basis asymmetry of the device, since the memory detectors show comparable rates for $X$- and $Z$-type errors. The origin of the separation therefore remains open.

Figure~\ref{Fig:QEC_experiment}(c) shows the logical error probability of the same patch versus the number of rounds for memories initialized in $|0\rangle_L$ and $|+\rangle_L$. In contrast to the two LR-QAOA frames, the two memory experiments show no comparable separation, differing by at most 25\% at any number of rounds. While the overall error level varies noticeably between sessions, by up to a factor of four at $R=3$, the dependence on $R$ remains similar. Normalizing each session to its value at $R=3$, all sessions follow a common curve to within about 10\%. The day-to-day variation therefore appears mainly as a change in the overall error scale rather than in how errors accumulate with the number of rounds.

Figure~\ref{Fig:QEC_experiment}(d) shows how the ranking agreement $\rho_s$ between the $X$-frame benchmark and the $|+\rangle_L$ memory depends on the benchmark shot budget. Larger $\rho_s$ indicates that the benchmark orders the patches more similarly to the memory experiment. We compare two protocols; a full sweep, where each depth $p$ is compared with the memory at the matched $R=p$, and a single-depth protocol using only $p=3$ against the memory at every $R$. The single-depth protocol performs better at every shot budget while requiring only one fifth of the cost, and it saturates quickly, reaching $\rho=0.81$ at 500 shots per circuit compared with $0.84$ at 1000. Each point is obtained by resampling, with replacement, from the 1000 experimental shots, which introduces additional sampling noise; the stars show $\rho$ from the full data without resampling ($0.86$ for $p=3$).

These results suggest a lightweight protocol for evaluating changes to the device, such as a new calibration procedure. Running $p=3$ with 500 shots in both frames on each candidate patch resolves $r_{\rm ovl}$ to about $\pm0.01$ after averaging over patches, sufficient to detect changes larger than the observed day-to-day drift.

\begin{figure}[!tbh]
\centering
\includegraphics[width=1\linewidth]{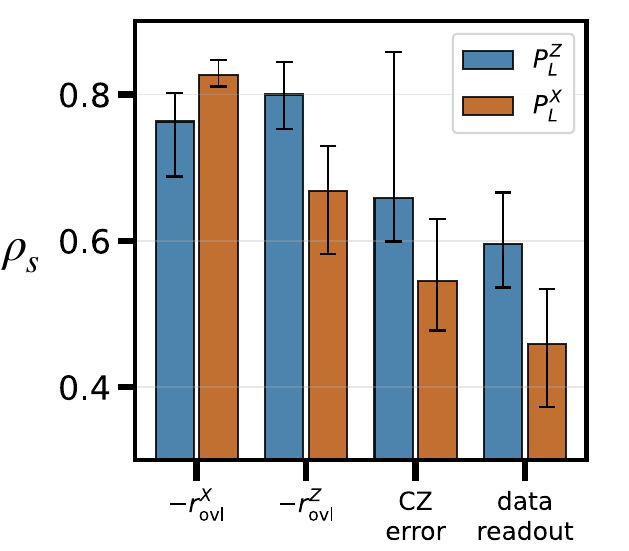}
\caption{\label{Fig:QEC_week} Spearman rank correlation $\rho_s$ between each predictor and the measured logical error probability of the $d=3$ surface-code memory on \texttt{ibm\_phoenix}, for memories initialized in $|0\rangle_L$ ($P_L^Z$) and $|+\rangle_L$ ($P_L^X$) after $R=3$ rounds. Each session ranks the same 11 patches; bars show the median over 12 sessions within one week, with error bars indicating the interquartile range. The LR-QAOA benchmark is evaluated at $p=3$ in the $Z$ and $X$ frames. Calibration predictors are taken from the device calibration at the time of each experiment and averaged over the corresponding patch; the mean CZ and mean data-qubit readout errors are shown as representative gate and readout metrics. All predictors are oriented so that larger values indicate a worse patch, such that larger $\rho_s$ corresponds to stronger agreement with the memory ranking.}
\end{figure}

Figure~\ref{Fig:QEC_week} summarizes how well each candidate predictor ranks the 11 patches by their logical error probability at $p=R=3$ over the week. In addition to the two LR-QAOA frames, we consider six calibration-based predictors measured at the time of each experiment: the mean and worst CZ error, mean single-qubit ($\sqrt{X}$) error, mean ancilla mid-circuit readout error, mean data-qubit readout error, and their sum over one syndrome round. The figure shows the mean CZ and data-readout errors as representative calibration metrics. 

The $X$-frame benchmark gives the strongest ranking agreement for the $|+\rangle_L$ memory ($\rho=0.83$), while the $Z$ frame gives the strongest agreement for $|0\rangle_L$ ($\rho=0.82$); the $X$ frame also remains competitive for $|0\rangle_L$ ($\rho=0.78$). For the $|+\rangle_L$ memory, the $X$ frame misplaces a patch by only 1.6 positions on average out of 11, compared with 2.1-2.7 positions for the calibration metrics and 3.6 for a random ordering. The advantage is smaller for $|0\rangle_L$, where the worst CZ error reaches a similar correlation ($\rho=0.78$). Most benchmark misrankings occur among the better patches, whose logical error probabilities differ by only a few percent.

This correspondence does not imply that LR-QAOA captures all aspects of QEC performance. The memory experiment interleaves $X$- and $Z$-type checks, builds detectors across rounds, and relies on decoding, while the benchmark evaluates commuting checks in a single frame. Our results are also limited to $d=3$ patches on one device. Within these limits, however, the agreement is quantitative and shows that LR-QAOA can serve as a practical indication for QEC performance, particularly for tracking improvements in the individual hardware and circuit-level components that enter a QEC protocol.

\section{Discussion and Conclusions}\label{Sec:Conclusions}

We have introduced a low-sampling protocol for characterizing quantum processors in the capacity of executing QEC primitives, mid-circuit measurement and QEC syndrome extraction structures. First, we run the LR-QAOA on a 1D-chain of qubits at different depths to see how MCM operations degrade the expected output of the algorithm. Then, we mimic the structure syndrome extraction of QEC codes using high-order Hamiltonians and MCM operations. In this way, the benchmark uses the same qubit footprint, connectivity, mid-circuit measurements, resets, and syndrome-extraction primitives required by the corresponding QEC implementation. Finally, we compare the LR-QAOA response with a logical surface-code memory experiment and find a correlation between the benchmark performance and the observed logical error probability.

Our main result is that MCM-based LR-QAOA provides a shot-efficient benchmark for evaluating the performance and capacity of QPUs to use QEC primitives. As hardware increasingly incorporates these capabilities, the benchmark can provide two complementary diagnostics. First, the homogeneity of the different hardware elements involved in implementing MCMs; and second, the response of the QEC structure as a whole. By comparing these results with the historical performance of previous device generations, one can track whether progress has been achieved in the design and fabrication of new QPUs. MCM-based LR-QAOA therefore provides a low-resource benchmark for an initial evaluation of QEC-relevant hardware performance.

Records of the maximum code distance and number of QEC instances that can be implemented on a given device would be useful for assessing the evolution of the broader ecosystem, particularly as many quantum-computing companies publish hardware roadmaps that include targets for logical-qubit capabilities. Organizations could apply a relatively simple benchmark of this kind and use the resulting historical record to assess whether successive hardware generations are progressing consistently with those roadmaps. At present, our results indicate that the performance degradation associated with MCM operations on IBM QPUs remains approximately an order of magnitude larger than that observed when executing the same protocol without MCM operations, whereas on Quantinuum QPUs the additional degradation associated with MCMs is comparable to that arising from two-qubit gate errors.

As a first application, we use LR-QAOA to assign an effective depolarizing rate to MCM implementations. This converts the observed decay of the LR-QAOA signal into a single perceived error rate, $\lambdaeff$, which can be compared across devices and circuit implementations. Applied to triplets, it shows the distribution of performance all over fix-layout QPUs. For the one-dimensional qubit-chain experiments, this analysis reveals behavior that is difficult to see in small systems alone. For example, on \texttt{quantinuum\_helios\_1}, the perceived error rate increases with chain length, indicating that the effective MCM cost is not fully captured by small-qubit experiments. Similar behavior is observed on the IBM devices, as the 1D-chain grows the perceived error also grows.

We also quantify the sampling cost required to interpret LR-QAOA data. Distinguishing a code-structured LR-QAOA distribution from the random-sampling limit requires only a small number of shots, and in the simulated SC and BB-qLDPC instances this cost does not grow with system size; it decreases as the distance or block length increases. This makes the random-baseline test a useful first diagnostic, before investing in larger shot budgets, one can ask whether the device still preserves algorithmic structure at all, or whether the output has already collapsed to the noise-dominated regime. Comparing two non-random devices is more demanding, but remains feasible with modest sampling. In that case the optimal LR-QAOA depth increases as the physical error rate decreases, because deeper circuits are needed to amplify the difference between two high-quality implementations. The resulting sampling cost is governed mainly by the relative gap between the perceived error rates.

Beyond the simulations, our experiments on \texttt{ibm\_phoenix} provide direct evidence of a correspondence between LR-QAOA performance and the outcome of SC memory experiments. Across different patches of the device, regions with better LR-QAOA performance generally exhibit lower logical error probabilities, with the strongest correlation reaching $|\rho_s|=0.88$ for $d=3$. This indicates that LR-QAOA experiments can capture hardware variations that are also relevant at the logical level, supporting their use as a preliminary benchmark for QEC performance. While this correspondence does not capture all aspects of a complete QEC implementation, but .

\section*{Data Availability}
All problem instances and code needed to reproduce the results of this work are available at \url{https://github.com/alejomonbar/qec-primitives-benchmark}.

\begin{acknowledgments}
\vspace{-10pt}

J. A. Montanez-Barrera acknowledge support from the project Jülich UNified Infrastructure for Quantum computing (JUNIQ) that has received funding from the German Federal Ministry of Research, Technology, and Space (BMFTR) and the Ministry of Culture and Science of the State of North Rhine-Westphalia (MKW-NRW). J. A. Montanez-Barrera acknowledges support from the project EPIQ funded by MKW-NRW. The authors gratefully acknowledge the Gauss Centre for Supercomputing e.V. (www.gauss-centre.eu) for funding this project by providing computing time on the GCS Supercomputer JUWELS Booster at Jülich Supercomputing Centre (JSC).

This research used resources of the Oak Ridge Leadership Computing Facility for the experiments on Quantinuum QPUs, which is a DOE Office of Science User Facility supported under Contract DE-AC05-00OR22725.

We acknowledge support from Amazon Web Services (AWS) through the provision of Amazon Braket credits used for quantum computing experiments on \texttt{iqm\_garnet} and \texttt{iqm\_emerald} .

We acknowledge the use of IBM Quantum Credits for this work. The views expressed are those of the authors, and do not reflect the official policy or position of IBM or the IBM Quantum team.
 
\end{acknowledgments}
\clearpage 
\bibliography{biblio.bib}

\clearpage  
\onecolumngrid  

\clearpage

\end{document}